# Enhancement of Luminescence of Quantum Emitters in the Epsilon-Near-Zero Waveguide

Jin-Kyu So[1*] and Nikolay I. Zheludev[1,2]

[1]The Photonics Institute & Centre for Disruptive Photonic Technologies, Nanyanag Technological University, Singapore 637371.

[2]Optoelectronics Research Centre & Centre for Photonic Metamaterials, University of Southampton, SO17 1BJ, UK.

*Correspondence to: jkso@ntu.edu.sg

**Abstract:** We report a resonant enhancement of luminescence intensity from an ensemble of CdS/ZnS quantum dots embedded in a nanoscale rectangular photonic waveguide operating at the epsilon-near-zero regime.

Enhancement of light emission from quantum emitters is one of the main goals of nanophotonics. The enhancement of a quantum system's spontaneous emission rate by its environment and, in particular, by confinement in a resonant cavity known as the Purcell effect[1], is widely used: multifold enhancements of the emission rate have been demonstrated in emitters embedded in plasmonic[2] and dielectric metamaterials[3]. Spectacular increase of spontaneous emission rate has been observed in quantum light emitters placed in nanoscale plasmonic resonators. An array of coupled plamonic-enhanced emitters can be forced into a collective mode of coherent emission by coupling between the resonators ("lasing spaser"[4]). It has been suggested that collective coherent emission of an ensemble of quantum emitters can also be achieved in epsilon-near-zero (ENZ) media where light experiences no spatial phase change and extremely large phase velocity[5]. Here, we report resonant enhancement of the intensity of emission of an ensemble of quantum emitters embedded in a nanoscale photonic waveguide operating at the epsilon-near-zero regime.

A rectangular waveguide is widely used as a microwave component which supports transverse electric (TE) and transverse magnetic (TM) modes for the wave transmission. When it is scaled down to nano-scale in the form of a dielectric core surrounded by metallic sidewalls, it supports the dominant quasi-TE mode which shows cutoff behavior and the position of this cutoff can be easily tuned with the refractive index, $n$, and width, $w$, of the dielectric core. It has been suggested that this type of waveguides can serve as an epsilon-near-zero medium near the cutoff frequency and exhibit the enhanced local density of optical states (LDOS) near such cutoff[6,7]. Enhanced luminescence of quantum emitters is expected when they are embedded in such waveguides whose cutoff is properly tuned to the emission wavelength of the emitters.

To construct a QDs-embedded waveguide as shown in Fig. 1a, we use Poly(methyl methacrylate) (PMMA) as a dielectric core with embedded CdS/ZnS quantum dots (emission wavelength ~ 630 nm, NN Labs). Figure 1b shows the effective index of quasi-TE mode for a PMMA core with $w =$ 150, 200, 300, 400 nm. The effective index is given by $k/k_0$ where $k$ and $k_0$ are the propagation constants of the guided mode and the electromagnetic wave in the free-space, respectively. A multi-layer film of silver, mixture of PMMA and QDs (~ 100-nm-thick, area density = 90 QDs/$\mu m^2$), and silver was deposited on a silicon substrate by thermal evaporation and spin-coating. The film was milled by focused ion beam to define the width of the waveguides. A subsequent deposition of silver film by thermal evaporation was followed to cover the exposed sidewalls. The structures were finalized with focused ion beam milling by carving the entrance and exit facets of the waveguides and forming two 45° mirrors for in- and out-coupling of pump laser and luminescence (see Fig. 1a).

The fabricated QDs-embedded waveguides were analyzed with a photoluminescence (PL) measurement setup where a 100x objective (NA=0.9x) was used to illuminate the waveguides with a 403 nm pump laser. The luminescence was collected with the same objective, coupled to an optical fiber and sent to either a single photon detector to produce a scanning PL intensity map or a spectrometer equipped with a thermoelectrically-cooled CCD. The sample was scanned by an xyz-piezo stage to obtain a luminescence intensity map as shown in the inset of Fig. 1a. The two bright spots in the scanning PL map (Fig. 1a inset) indicate the efficient coupling of the pump laser into the waveguide via the two 45° mirrors. To measure the PL spectra of the waveguides, the excitation was placed on one of these bright spots after taking a scanning PL intensity map around each waveguide.

By skipping the last silver deposition, a control waveguide was prepared where every feature of QDs-embedded waveguides remains the same except the sidewalls being absent. This ensures the control waveguide does not show any cutoff or strong resonance behavior near the spectral region of our interest, i.e. QD emission wavelength, ~ 630 nm. As a result, the luminescence intensity from QDs in such control waveguides monotonically increases with the width of the waveguide (Fig. 2a), which is attributed to the increase in the number of QDs embedded in the PMMA core as the width of a 2.4-µm-long waveguide is increased from 100 to 180 nm.

However, when the PMMA core is fully surrounded with a metal film, the luminescence from QDs is a strong function of the waveguide width. Figure 2b shows the PL spectra from QDs-embedded waveguides where the spectra are normalized by matching the background luminescence level at 570 nm to unity. The luminescence from QDs in a 100-nm-wide waveguide is strongly suppressed and the spectrum is identical to that from unstructured surface on the same sample. However, as the width is gradually increased, the luminescence shows a sudden jump in the peak intensity for $w = 160$ nm (Fig. 2b). The observation of suppressed QD luminescence for narrow waveguide width, $w = 100 – 150$ nm, and the enhancement for $w = 160$ nm can be understood as a result of the change in local density of optical states (due to ENZ behavior) at QD emission wavelength as a function of the waveguide width. With the further increase of the waveguide width, QD luminescence undergoes a series of suppression and enhancement (due to Fabry-Perot resonances) along the waveguide axis).

In conclusion, we have shown that nanoscale rectangular waveguides can be used to control the light emission from quantum emitters with a specific interest in its epsilon-near-zero behavior near its cutoff. As an exemplary system, we have implemented a QDs-embedded waveguide with a

PMMA core which showed the suppression and enhancement of the QD luminescence corresponding to the change in local density of optical states of the waveguide.

This work was supported by the Singapore ASTAR QTE program (No. SERC A1685b0005), the Singapore Ministry of Education (No. MOE2016-T3-1-006 (S)), and the UK's Engineering and Physical Sciences Research Council (Grant No. EP/M009122/1).

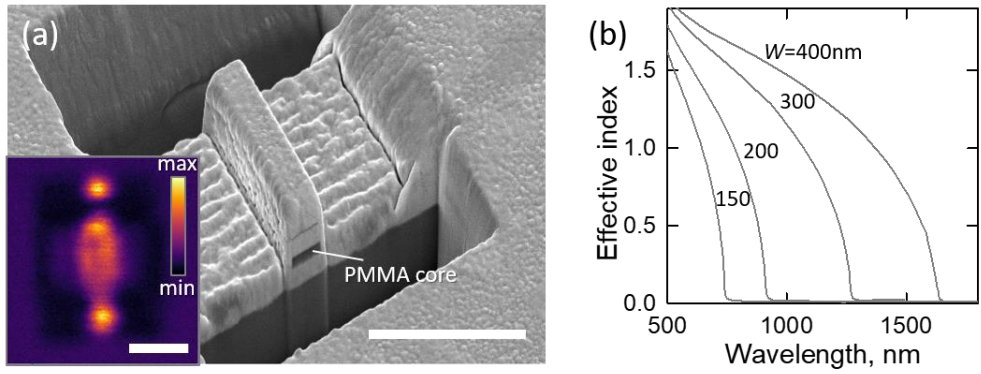

Figure 1. (a) Scanning electron microscope image and (inset) scanning PL intensity map of a 400-nm-wide and 2.4-μm-long QD-embedded waveguide with two 45° reflectors. The top and bottom bright spots in the PL map indicate the positions on the 45° reflectors for efficient in- and out-coupling of pump and emitted light. Scale bars are 2 μm. (b) effective refractive index of quasi-TE mode in waveguides with $w$ = 150, 200, 300, 400 nm.

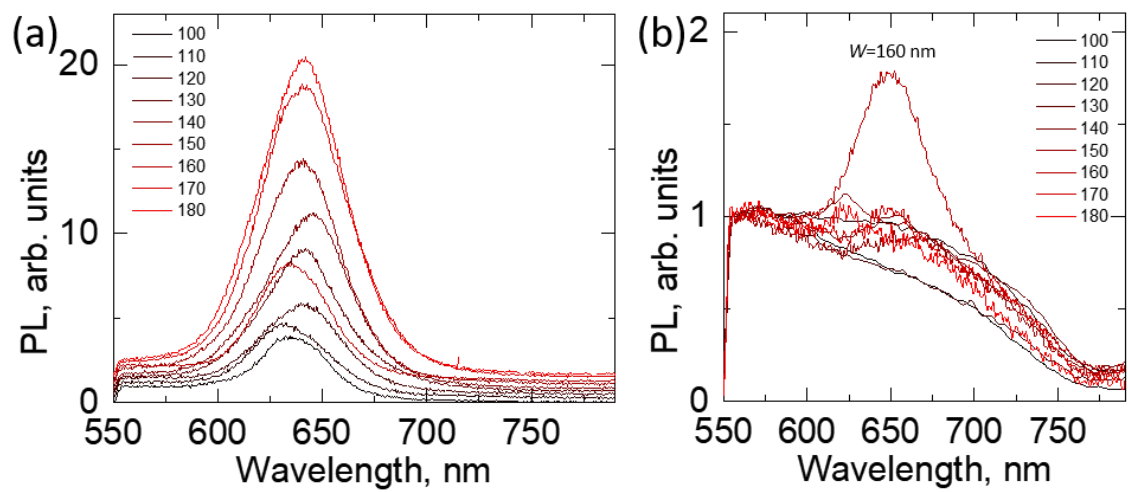

Figure 2. (a) Measured PL spectra from control waveguides (without sidewalls) with waveguide width, $w$ = 100 – 180 nm. (b) Measured PL spectra from QDs-embedded waveguides (with sidewalls) with $w$ = 100 – 180 nm.